\documentclass[final]{IEEEtran}

\usepackage[english,brazilian]{babel} 
\usepackage[utf8]{inputenc}
\usepackage[T1]{fontenc}
\usepackage{listings}
\usepackage{graphicx}
\graphicspath{{images/}}
\usepackage{wrapfig}
\usepackage{rotating}
\usepackage{algorithm}
\usepackage[noend]{algpseudocode}

\usepackage{cite}

\algdef{SE}[SUBALG]{Indent}{EndIndent}{}{\algorithmicend\ }%
\algtext*{Indent}
\algtext*{EndIndent}

\addto\captionsbrazilian{%
}

\begin{document}

\title{A Solution for Controlling and Managing User Profiles based on Data Privacy for IoT Applications}

\author{
Valderi R. Q. Leithardt\IEEEauthorrefmark{1}\IEEEauthorrefmark{2}, \IEEEauthorblockN{Douglas A. dos
Santos\IEEEauthorrefmark{1}, 
Luis A. Silva\IEEEauthorrefmark{1},
Felipe Viel\IEEEauthorrefmark{1},\\   Cesar A. Zeferino\IEEEauthorrefmark{1},  Jorge Sa Silva\IEEEauthorrefmark{2}
}

\IEEEauthorblockA{\IEEEauthorrefmark{1}Laboratory of Embedded and Distributed Systems, University of Vale do Itajai (UNIVALI), Brazil, 88302-901}

\IEEEauthorblockA{\IEEEauthorrefmark{2}Department of Computer Engineering, University of Coimbra, Portugal, 3000-370\\
Email: valderi@univali.br, \{douglasas, luis.silva, felipeviel\}@edu.univali.br, zeferino@univali.br, sasilva@dei.uc.pt}
}

\maketitle
\selectlanguage{english}
\begin{abstract}
IoT is an emerging area in which we expect to have billions of devices connected to the Internet by 2020. IoT applications can offer many benefits to environments, society and the economy through the interconnection and cooperation of smart objects. However, there are many privacy challenges, such as authentication, authorization, and confidentiality of personal data. With this in mind, we have developed a solution for managing user profiles based on privacy and evolution. For that, we define the criteria and characteristics for each environment. The contribution of the work also applies to the evolution of the UbiPri Middleware, with a module implemented and tested according to the rules of environments so that they can modify the user's profile over time according to their frequency in the environment. The modified profile can be raised, reduced and blocked. This implementation has been validated using scripts that perform probabilistic simulation and user authentication. From the rules assigned to the environments, it was possible to perceive the high adaptability of the implementation, and it can be easily adjusted to any IoT environment that wants to treat the authentication and privacy of environments and users. The results obtained were satisfactory and could be useful for both the evolution of the UbiPri Middleware and other related works.
\end{abstract}

\IEEEoverridecommandlockouts
\IEEEpeerreviewmaketitle

\selectlanguage{brazilian}

\section{Introdução}

A IoT pode ser vista de diversas perspectivas e formas, dentre as quais, destacam-se as abordagens orientadas à Internet, às coisas e ao conhecimento. Quanto à forma, basicamente, a IoT pode ser dividida em dar suporte a humanos e dar suporte a aplicações industriais, descreve  \cite{Kubitza2016}. Essas duas formas recebem os nomes, respectivamente, de Internet Humana das Coisas (HIoT – Human Internet of Things) e Internet Industrial das Coisas (IIoT – Industrial Internet of Things). Apesar disso, a IoT se define em função da aplicação à qual está se inserindo e ao objetivo que se quer alcançar. 


Os três elementos principais que podem ser considerados em um sistema IoT são o hardware, o middleware e a apresentação. O hardware é o responsável por fazer a aquisição de dados por meio de sensores, dar suporte ao tratamento de dados com unidades de processamento e atuar, quando necessário, por meio de atuadores. O middleware é responsável por dar suporte ao desenvolvimento rápido de aplicações que necessitam armazenar, processar e analisar dados, abstraindo o sistema operacional e o hardware. O middleware também é responsável por permitir o acesso de diferentes formas e por diferentes usuários (sistemas computacionais e humanos) ao conhecimento gerado \cite{wortmann2015internet, gubbi2013internet}. 
As características principais das aplicações IoT se resumem, conforme \cite{razzaque2016middleware}: em (\textit{i}) diversidade, pois as diferentes aplicações apresentam requisitos diversos e provavelmente necessitam de arquiteturas diferentes; (\textit{ii}) tempo real. Para realizar as ações necessárias de acordo com os requisitos temporais da aplicação; (\textit{iii}) segurança e privacidade, para proteger as aplicações, as redes e os usuários; e (\textit{iv}) modelo de serviços, sendo o modelo XaaS (Everything-as-a-Service) bastante eficiente, escalável e de fácil uso, conforme  \cite{banerjee2011everything}.  Consequentemente, pesquisas descritas em \cite{li:18}, entre outras relacionadas a IoT afirmam que a segurança e privacidade como sendo um grande desafio e problema de pesquisa   fundamental para permitir que a IoT seja segura, acessível e adotada em larga escala. 
Dentro deste contexto, com o objetivo de proporcionar maior segurança e privacidade aos ambientes que integram sistemas, aplicativos e demais tecnologias IoT, este trabalho apresenta contribuição relacionada a definição de métricas, parâmetros e critérios de hierarquia de perfis para o gerenciamento de privacidade. Para tanto,  foi desenvolvido um módulo denominado PriPro (\textit{Privacy Profiles}) do modelo de controle de privacidade para ambientes pervasivos/ubíquos (UbiPri) proposto por \cite{leithardt:13}. 
A outra contribuição relevante deste trabalho foi obtida nos testes realizados, obtendo a  possibilidade de atribuir ou diminuir requisitos ao usuário de acordo com o tempo estabelecido de acordo com a identificação e permanência no ambiente localizado com uso de dispositivos. Para tanto, os ambientes foram definidos em 5 categorias, e os níveis de usuários para acessar e utilizar também em 5 categorias. Com isso, identificamos e atribuímos diferentes níveis de privacidade de acordo com a hierarquia definida individualmente. 
Para tanto, estruturamos o trabalho para um melhor entendimento em sete seções. A Seção \ref{sec:seguranca_privacidade} apresenta conceitos base sobre segurança e privacidade, enquanto a Seção \ref{sec:trabalhos_relacionados} discute os trabalhos relacionados. A Seção \ref{sec:middleware} descreve o middleware de referência do trabalho, enquanto a Seção \ref{sec:modelagem} apresenta a modelagem da solução proposta. A Seção \ref{sec:experimento} descreve o cenário de teste empregado, os testes realizados e os resultados obtidos. A Seção \ref{sec:prototipo} ilustra o protótipo de sistema embarcado implementado para validação da solução proposta. Por fim, a Seção \ref{sec:conclusoes} apresenta as conclusões.

\section{Serviços de Segurança e Privacidade}
\label{sec:seguranca_privacidade}

Os princípios básicos que devem estar presentes na segurança e privacidade de informações e redes de computadores são: (\textit{i}) confiabilidade; (\textit{ii}) integridade; (\textit{iii}) disponibilidade; (\textit{iv}) autenticação; e (\textit{v}) autorização. Esses princípios, diferentemente de aplicações convencionais da internet, têm um grau maior de importância em IoT, dado que esses sistemas têm um impacto econômico maior mundialmente \cite{kraijak2015survey}. As questões de segurança e privacidade recebem maior atenção, além do impacto econômico, porque os sistemas de IoT, invisíveis a usuários, têm uma aquisição de dados densa e de forma pervasiva, além do processamento e disseminação de dados de usuários. Assim, o desprezo por essas questões citadas conduz a sérias consequências comerciais, civis e jurídicas \cite{ziegeldorf2014privacy}.

Uma alta heterogeneidade de sistemas e dispositivos que estão integrados, em larga escala, e as restrições de recursos impostas por esses dispositivos, como memória e processamento, torna a implementação de técnicas de segurança e privacidade já consolidadas para aplicações convencionais para internet de difícil integração \cite{sicari2015security}. Segundo \cite{zhang2014iot}, em segurança, os fatores de impacto podem ser divido em duas categorias: a diversidade das coisas e a comunicação entre as coisas. Essa divisão em duas categorias permite avaliar as questões e diferentes problemas de segurança com maior foco em cada categoria. Essa categoria é considerada mais complexa que as encontradas até então em outras arquiteturas de sistemas. Já a categoria comunicação entre as coisas, enfrenta a heterogeneidade entre meios de comunicação que requerem diferentes níveis de segurança, os quais não se pode negligenciar para não comprometer as coisas \cite{RUBIO2017205}. Nessa categoria, há um esforço de segurança sobre as estruturas de dados e protocolos usados.

Em \cite{kraijak2015survey}, os autores descrevem que as principais pesquisas relacionadas à segurança focam em protocolos de comunicação para M2M. Quanto à vulnerabilidades do sistema, o autor divide o sistema IoT em partes: (\textit{i}) sensores e equipamentos de \textit{front-end}; (\textit{ii}) redes; e (\textit{iii}) \textit{back-end}. Em (\textit{iv}), a segurança é colocada em risco quando exploradas vulnerabilidades no envio de dados amostrados, podendo haver desvio das informações com a reprogramação de dispositivos. Em (\textit{ii}), como são responsáveis pelo tráfego de todos os dados da comunicação M2M, assim, podem sofrer com ataques relacionados a QoS, como negação de serviço. Por fim, em (\textit{iii}), há a parte mais importante dos sistemas IoT, requisitando níveis mais elevados de segurança e eficiente análise de dados provenientes de sensores para garantir processamento em tempo real pelo sistema. Quanto às soluções para segurança, os autores colocam que essas podem ser implementadas na identificação, autenticação e controle de recursos. A identificação pode ser realizada com Domain Name Service Security Extension (DNSSEC). A autenticação com criptosistemas como Datagram Transport Layer Security (DTLS), Key Management System (KMS) e Public Key Infrastructure (PKI). O controle de recursos com Owner-extended Role-Based Access Control (OxRBAC) e Capability Based Access Control (CapBAC) \cite{sicari2015security}.




A privacidade pode ser vista como o correto tratamento de diferentes tipos de dados que, quando não realizado, por exemplo, com o uso de Digital Right Management (DRM), pode sofrer com ataques que armazenam e enviam os dados a terceiros \cite{kraijak2015survey}. Com o intuíto de tratar a privacidade em autenticação como contramedidas para aumentar os níveis de privacidade do sistema IoT, são utilizados serviços como DNSSEC para localização e identificação,  Anonymous Access Credential AAC e ePASS, baseado em Attribute-Base Signature (ABS) \cite{sicari2015security}. Além disso, a segurança e a privacidade ainda podem ser reforçadas com o uso de middlewares para o devido fim, o qual permite o uso de diferentes níveis e protocolos voltados a requisitos. Exemplos de middlewares voltados a segurança e privacidade são VIRTUS, Otsopack, NAPS, OneM2M \cite{sicari2015security}. Outros exemplos de trabalhos que também tratam privacidade, e auto grau de precisão de classificação de dados como por exemplo \cite{ILATransactions}, mas trabalhos encontrados na literatura que estão mais relacionados ao controle de acesso aos ambientes, são apresentados na Seção III.

\section{Trabalhos Relacionados}
\label{sec:trabalhos_relacionados}

Os autores em \cite{hussein:17} apresentaram a proposta de um \textit{framework} de controle de acesso orientado a comunidades chamada COCapBAC (Community Capability-Based Access Control). Nele é proposto que dispositivos inteligentes dentro de uma mesma comunidade possuam as mesmas restrições de acesso. Sendo que, quando um dos dispositivos da comunidade realizar uma autenticação, os outros precisam apenas se comunicar com o dispositivo de controle de acesso (AC) que se localiza dentro da rede interna para buscar as permissões do usuário, aumentando o desempenho da rede para usuários recorrentes. Também foi implementado um protótipo, no qual foi utilizado o microcontrolador ATMEGA328 e a identificação das pessoas é feita com smartphones. A abordagem de \cite{hussein:17} se assemelha muito a deste trabalho, porém, o presente trabalho apresenta vantagens, pois ele abre mais possibilidades de acesso aos dispositivos devido à definição de perfis. Além disso, o presente trabalho também implementou um protótipo, porém foi utilizado o microcontrolador ATMEGA328P, que consome menos energia, e a identificação das pessoas é feita com tags RFID (do inglês, Radio Frequency IDentification).

No trabalho \cite{gusmeroli:13}, foi desenvolvido o sistema CapBAC (Capability Based Access Control), o qual é utilizado como o controlador de acesso ao middleware de automação.
Uma desvantagem desse sistema em relação ao presente trabalho, é que ele não é automático, os administradores do ambiente devem informar manualmente os usuários que terão acesso.

A solução apresentada em \cite{bernabe:16}, é um mecanismo de controle de acesso confiável para IoT denominado TACIoT (Trust-Aware Control Mechanism for IoT) para lidar com a imprecisão associada com as informações do contexto em cenários pervasivos. O CapBAC propõe um novo modelo de confiança baseado na lógica fuzzy, que é o pilar do TACIoT. Esse modelo segue a abordagem multidimensional para possibilitar uma computação de valor confiável e preciso para dispositivos IoT \cite{Rodrigues:17}. Diferente de modelos anteriores que apenas consideram a reputação e o \textit{feedback}, utiliza os aspectos de segurança e os fatores sociais entre dispositivos IoT, que são utilizados por objetos inteligentes para conduzir lógica do controle de acesso.

A Tabela \ref{tab:comparacao} apresenta a comparação do presente trabalho com os trabalhos analisados. A tabela identifica se os trabalhos trataram a privacidade de dados, utilizaram perfis de usuários para tratar a privacidade, trataram privacidade e segurança de ambientes IoT, implementaram uma recompensa ou punição de acordo com a permanência do usuário no ambiente, desenvolveram um sistema genérico que pode ser implementado em ambientes com diferentes características e se propuseram um sistema automático para decidir o acesso do usuário.

\begin{table}[htbp]
\centering
\caption{Análise comparativa}
\label{tab:comparacao}
\begin{tabular}{lcccccc}
\hline
  & \cite{sicari2015security} 
  & \cite{hussein:17} 
  & \cite{gusmeroli:13} 
  & \cite{bernabe:16} 
  & \cite{santos:17} 
  & \parbox{1.2cm}{\vspace{0.05cm}Este\\Trabalho\vspace{0.05cm}} 
  \\\hline
	Privacidade de Dados		& \textbullet & & & & \textbullet & \textbullet \\
    Perfis de Usuários			& & & & & \textbullet & \textbullet \\
  	\parbox{2.5cm}{\vspace{0.05cm}Privacidade de\\Ambientes\vspace{0.05cm}}	& \textbullet & \textbullet & \textbullet & \textbullet & \textbullet & \textbullet \\
  	Evolução de perfis			& & & & & \textbullet & \textbullet \\
  	Sistema Genérico			& \textbullet & \textbullet & & \textbullet & & \textbullet \\
  	Sistema Automático			& & \textbullet & & \textbullet & \textbullet & \textbullet \\
  	Sistema Embarcado			& & \textbullet & & & & \textbullet \\\hline
  
\end{tabular}
\end{table}

Como pode ser observado na tabela I, o presente trabalho destaca-se por tratar a privacidade de ambientes IoT com um sistema genérico e automático, contribuindo com a implementação para a evolução dos perfis. Este trabalho também simula a evolução de confiança nos usuários em ambientes IoT, dando a elas mais ou menos permissões na medida em que utilizam o ambiente (Evolução de perfis). Além disto, este trabalho apresenta a implementação de um protótipo que utiliza este sistema, voltado para IoT.

Este artigo é uma evolução de um trabalho anterior desenvolvido por \cite{santos:17}, no qual foi proposta uma estratégia para definição de perfis de usuários. O diferencial em relação ao referido trabalho reside na implementação da evolução dos perfis dos usuários em relação à permanência no ambiente e no desenvolvimento de um protótipo de um sistema embarcado, para o controle de acesso de pessoas. Na evolução de perfis, o perfil pode tanto aumentar ou diminuir na ordem hierárquica, automaticamente de acordo com as definições e regras. Adicionalmente, o trabalho também se fundamentou no modelo taxonômico de \cite{leithardt:13}, que apresentou módulos necessários para o tratamento da privacidade nos ambientes. Os trabalhos em questão são detalhados na seção a seguir.

\section{Middleware de Referência}
\label{sec:middleware}

Esse trabalho é fundamentado no Middleware UbiPri (Ubiquitous Privacy) \cite{leithardt:13}, em continuidade aos estudos realizados em \cite{santos:17}. O middleware UbiPri segue um modelo genérico, o qual contém diversos componentes para controlar ambientes pervasivos e ubíquos. 
As funções do software de rastreamento do usuário, tomada de decisão e adaptações para suportar o gerenciamento de privacidade estão definidas conforme \cite{leithardt:18}.
O módulo desenvolvido nesse trabalho é o PRIPRO (PRIvacy PROfiles), que, de acordo com \cite{leithardt:13}, realiza o controle de transações no gerenciamento do perfil do usuário. Seu principal objetivo é controlar a informação, definida previamente por um motor de busca que possuía o propósito de distribuir e direcionar toda a informação sintetizada para os módulos seguintes, a fim de adaptar apropriadamente para a privacidade individual com base no perfil individual.

Para tanto, foi desenvolvido o modulo de controle de perfis (PRIPRO) para o middleware UbiPri. Esse módulo utiliza serviços web com o modelo arquitetural REST (REpresentational State Transfer). Esse middleware é responsável por definir, para cada usuário, limitações no uso de objetos inteligentes presentes nos ambientes.

\begin{figure}[htbp]
	\centering
    \includegraphics[scale=.35]{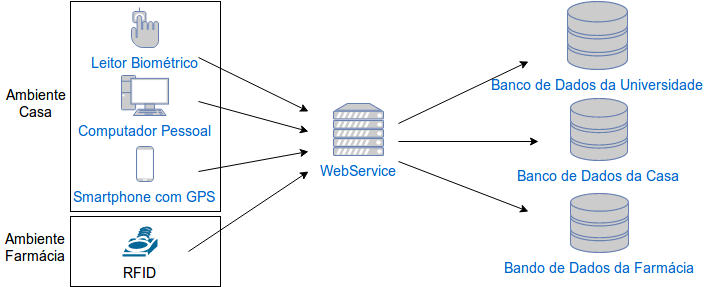}
    \caption{Arquitetura de utilização do serviço Web \cite{santos:17}.}
    \label{fig:arc_servico}
\end{figure}

O desenvolvimento do módulo PRIPRO utilizou serviços web com o modelo arquitetural REST e recebe requisições de dispositivos que desejam autenticar usuários no ambiente. Esses dispositivos podem ser leitores biométricos, leitores de RFID (Radio Frequency IDentification), smartphones, entre outros. O serviço web informa ao dispositivo o perfil do usuário, dados do ambiente, recursos e serviços disponíveis utilizando JSON (JavaScript Object Notation). O perfil do usuário é definido a partir dos dados que estão disponíveis para o módulo. Esses dados são informações do usuário enviadas pela requisição e obtidos dos bancos de dados específicos da esfera em que ele se localiza. A Fig. \ref{fig:arc_servico} exemplifica a utilização dos bancos de dados específicos. Neste exemplo, o serviço web tem acesso ao banco de dados específico de três esferas: (\textit{i}) Universidade; (\textit{ii}) Casa; e (\textit{iii}) Farmácia. No caso de uma requisição ser realizada do ambiente Farmácia, o serviço web deve identificar a esfera de onde originou a requisição e buscar os dados do banco específico dessa esfera. Isso é feito para garantir que os dados de uma esfera não influenciem nas outras.

Cada esfera possui as suas regras e particularidades, fazendo com que cada uma delas esteja em um contexto totalmente diferente. Para que uma requisição ao sistema seja válida, há a necessidade de indicar certos parâmetros que permitam a busca dos dados do usuário. Visando a simplicidade da requisição, foram definidos como parâmetros: (\textit{i}) o identificador do dispositivo do usuário; (\textit{ii}) o que o usuário pretende fazer (entrar ou sair do ambiente); e (\textit{iii}) o identificador do dispositivo autenticador. Desse modo, os dados necessários nos dispositivos são mínimos, simplificando a implementação.

\section{Modelagem da Evolução dos Perfis}
\label{sec:modelagem}

A partir dos resultados preliminares  foram desenvolvidas implementações adicionais relacionadas à evolução do perfil do usuário. Para tanto, foi necessário estabelecer os períodos que seriam tratados pela esfera, ou seja, os períodos para os quais é calculada e salva a frequência de cada um dos usuários da esfera. Esses períodos são definidos neste artigo com períodos de valência (períodos que, juntos, serão utilizados para calcular a frequência do usuário e definir o seu perfil).

A tabela \textit{Period} possui todas as opções de períodos que podem ser utilizados para definir os perfis dos usuários. Na tabela \textit{ExpectedFrequency}, são escolhidas, para cada esfera, os períodos que cada esfera irá tratar. Também é definido o tempo em horas que espera-se que os usuários permaneçam no ambiente. A tabela \textit{Frequency} guarda as frequências calculadas dos usuários em cada ambiente, otimizando o sistema de modo que, quando o usuário se autenticar nos ambientes, o sistema não precise calcular a sua frequência, acelerando sua resposta.


Em (\ref{eq:finf}), é calculada a frequência do usuário no ambiente. Quando o período e/ou turno é identificado como devendo ser inferior, será calculado o período de valência mais inferior (\textit{f\textsubscript{inf}}), mantendo, assim, a mesma hierarquia e definições inferiores para aquele ambiente e turno identificados. Considerando um cenário ideal, um usuário gera uma autenticação de entrada e uma de saída para cada vez que ele ingressa ou se retira do ambiente. A equação consiste em somar os intervalos de tempo entre a entrada e a saída (\textit{t\textsubscript{entrada}} e \textit{t\textsubscript{saida}}, respectivamente) e dividir pelo tempo que o ambiente espera que o usuário permaneça nele (\textit{t\textsubscript{esp}}), obedecendo a premissa deste trabalho de que o ambiente que define as suas regras.

\begin{equation}
\label{eq:finf}
f_{inf} = 1/(t_{esp}) \sum (t_{i saida} - t_{i entrada})
\end{equation}

Quando for necessário calcular a frequência do usuário no ambiente para períodos de valência superiores, utiliza-se (\ref{eq:fsup}). Essa equação consiste em calcular a média das frequências(\textit{f\textsubscript{sup}} -- frequência no período superior, o qual deseja ser descoberto), dentro do período de tempo, do período de valência imediatamente inferior (\textit{f\textsubscript{inf}} -- frequência no período imediatamente inferior), que serão divididas pela quantidade de períodos de valência imediatamente inferiores que cabem no superior (\textit{n}).

\begin{equation}
\label{eq:fsup}
f_{sup} = 1/n \sum f_{inf}
\end{equation}

O cálculo da frequência dos usuários é realizado toda vez que o período de valência mais inferior da esfera é concluído. Ele consiste em somar o tempo que o usuário permaneceu no ambiente e dividir pela quantidade de horas que o ambiente espera que ele permaneça, ou seja, o ambiente define as suas regras de acesso (\ref{eq:finf}). Porém, quando constatado que períodos superiores já possuem seu intervalo de tempo limite expirado, são calculadas as suas respectivas frequências por meio da média dos períodos imediatamente inferiores desde a última atualização do período desejado (\ref{eq:fsup}). Com a frequência calculada, os usuários evoluem seus perfis de acordo com as regras definidas para cada esfera.

\begin{figure}[htbp]
	\centering
    \includegraphics[scale=.7]{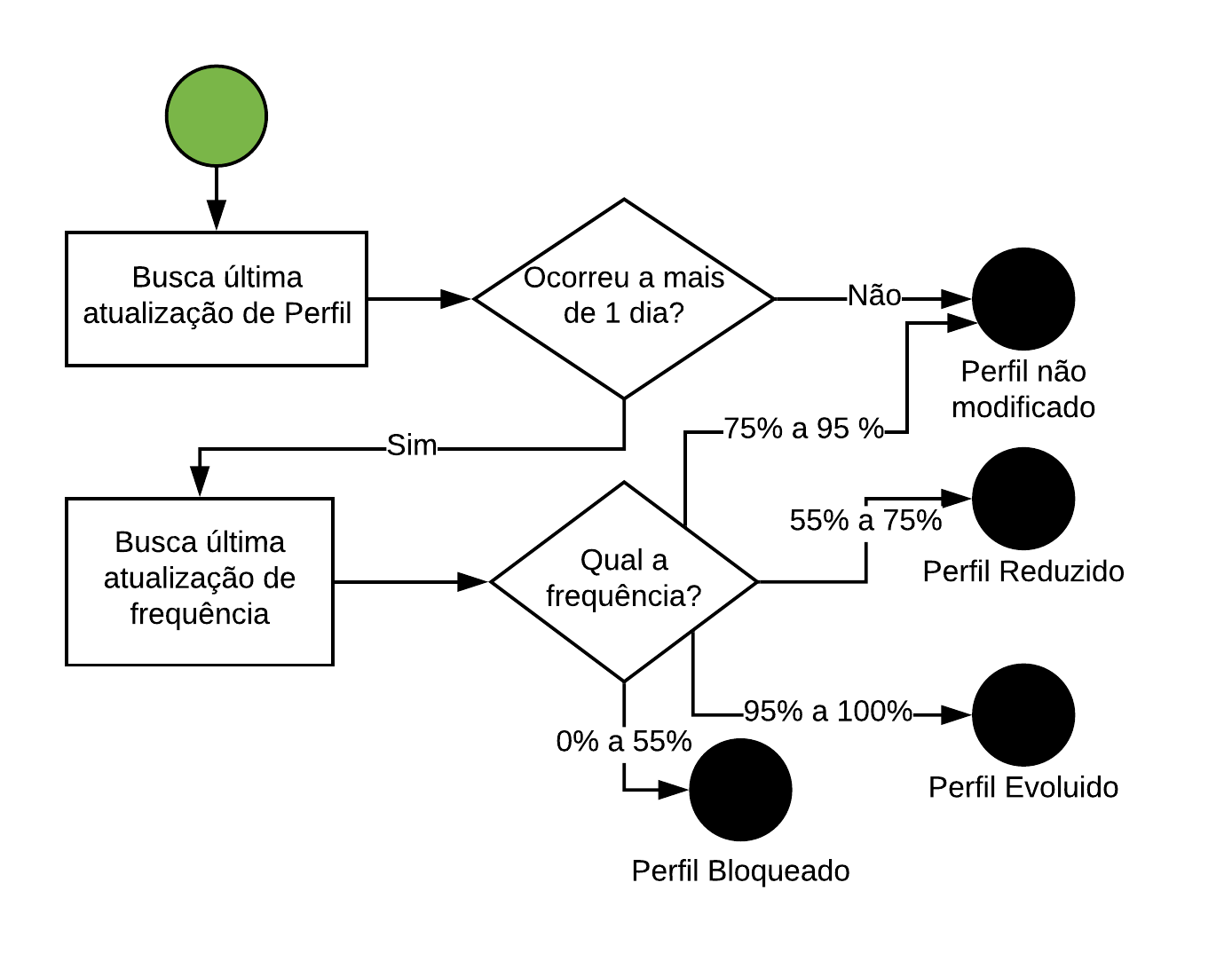}
    \caption{Fluxograma de evolução do perfil do usuário teste.}
    \label{fig:atividade}
\end{figure}

A frequência de atualização dos perfis é definida por meio do tempo do período mais inferior da esfera, definindo de quanto em quanto tempo que o usuário poderá ter seu perfil evoluído. Na implementação realizada no trabalho anterior \cite{santos:17}, foi definida a técnica para implementação de um sistema genérico de definição de perfis, na qual deveria ser realizada uma requisição ao banco de dados específico da esfera. Nesse trabalho, o perfil do usuário é definido a partir do perfil que o usuário obteve no último acesso. Porém, quando o usuário ainda não se autenticou no ambiente, ou seja, não é conhecido pelo sistema, uma requisição ainda deve ser feita para o banco de dados específico da esfera para obter um perfil base. Depois de obtido o perfil base, para todas as autenticações seguintes, são feitas as verificações mostradas no fluxograma ilustrado na Fig. \ref{fig:atividade}.

Toda vez que o usuário se autenticar no ambiente, é feita a análise descrita no fluxograma. Primeiro, é buscada a última atualização do perfil do usuário (última autenticação no ambiente) e verificado se ocorreu a um tempo maior do que o período de valência mais inferior. Se não ocorrer, o perfil não é modificado. Se ocorrer, o sistema busca a última atualização de frequência do usuário na tabela \textit{Frequency}. A partir do valor da frequência, são aplicadas as regras do ambiente para definir se o perfil do usuário será modificado (reduzido, evoluído ou bloqueado) ou se ele não será modificado. Com o objetivo de realizar testes e validar a implementação, a seção a seguir apresenta o cenário de testes utilizado. 

\section{Protótipo Desenvolvido}
\label{sec:prototipo}

Para os testes e validação, foi desenvolvido um protótipo de sistema embarcado para o controle de acesso das pessoas, sendo este um dispositivo autenticador. O protótipo implementa o controle do acesso com tags RFID e definições de comunicação conforme descrito em \cite{7460923}. O dispositivo faz uma requisição para o Serviço Web informando o ID do dispositivo autenticador e do dispositivo do usuário (tag RFID), que analisa os dados da pessoa e do ambiente e informa perfil definido de acordo com os critérios e parâmetros.

O protótipo utiliza um microcontrolador ATMEGA328P de baixo consumo de energia, uma \textit{shield} Ethernet W5100 para conexão à Internet, um módulo MFRC522 para leitura das tags RFID e um LCD 20$\times$4 para exibição de mensagens, conforme ilustrado na Fig.{fig:prototipo}. O microcontrolador utiliza o protocolo SPI (do inglês, Serial Peripheral Interface) para comunicação com o W5100 e com o MFRC522, e o protocolo I2C (do inglês, Inter-Integrated Circuit) para comunicação com o LCD. A comunicação para a requisição dos dados é realizada com o uso do protocolo HTTP, sendo que as informações são transferidas em formato JSON.

\begin{figure}[htbp]
	\centering
    \includegraphics[scale=0.37, angle=90]{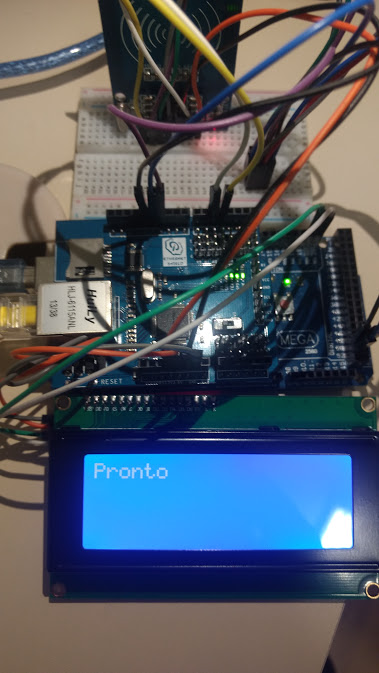}
    \caption{Protótipo do dispositivo de controle de acesso.}
    \label{fig:prototipo}
\end{figure}

O protótipo desenvolvido funcionou de acordo com as definições, apresentando, ao usuário os recursos e serviços a ele disponibilizados no ambiente. O consumo de corrente foi de, aproximadamente, 300 mA, principalmente para alimentação do LCD. As requisições para o servidor não demoraram muito tempo, porém, em trabalhos futuros, pretende-se utilizar técnicas de Fog Computing em ambientes que necessitam uma alta disponibilidade.

\section{Resultados obtidos}
\label{sec:experimento}

\subsection{Cenário de Testes}
\label{subsec:cenario_testes}

A implementação realizada se concentrou no cenário de testes de ambientes de uma universidade. O cenário define que o período de valência mais inferior é o dia, o que significa que os usuários podem ter seu perfil modificado uma vez por dia, impedindo que o mesmo usuário, em um mesmo dia, tenha dois perfis diferentes. Além disso, de acordo com a frequência, o perfil do usuário será: (\textit{i}) bloqueado, se possuir uma frequência menor do que 55\%; (\textit{ii}) reduzido, se possuir uma frequência entre 55\% e 75\%; (\textit{iii}) mantido, se possuir uma frequência entre 75\% e 95\%; e (\textit{iv}) evoluído, se possuir uma frequência maior que 95\%, conforme observado na Fig. \ref{fig:atividade} demonstra a modelagem utilizando um fluxograma do código de atualização de perfis. 

O cenário de ambiente testado foi um laboratório, no qual é definido que o tempo esperado para que o usuário permaneça no ambiente é de 8 horas diárias. Assim, sua frequência esperada (\textit{f\textsubscript{esp}}), aplicada na equação para o cálculo da frequência do usuário (\ref{eq:finf}), é 8 horas. Neste cenário, os períodos de valência utilizados, da ordem do mais abrangente até o menos abrangente, são: \textit{Semestre}, \textit{Mês}, \textit{Semana} e \textit{Dia}. 
Nos testes realizados foi possível evidenciar que a frequência que o usuário teve no dia anterior pode modificar o seu perfil no dia atual. O estado \textit{Frequência}, na parte inferior da figura, são mostradas as modificações da frequência do usuário de acordo com os dias. O estado \textit{Perfil} significa o perfil do usuário, que é mantido durante um dia e pode sofrer modificações a cada dia. As setas indicam o valor da frequência que fez com que o estado \textit{Perfil} fosse modificado.


Observou-se também que o usuário obteve o perfil básico devido à requisição realizada para esfera específica. Inicialmente, ele mantém o mesmo perfil, pois está frequentando o ambiente dentro dos parâmetros aceitáveis. Porém, nos dias 3 e 5, ele frequentou o ambiente durante um período igual ou maior do que 7,6 horas, fazendo com que, nos dias 4 e 6, seu perfil evoluísse de básico para avançado e de avançado para administrador, respectivamente. Nos dias 9 e 10, o usuário frequentou o ambiente durante um período entre 4.4 e 6 horas diárias, causando uma diminuição de perfil em 2 dias consecutivos (10 e 11). Nos dias 11 e 14, ele permaneceu dentro dos parâmetros aceitáveis, mantendo assim o perfil sem evoluções, até que, no dia 15, ele permaneceu por um período menor do que 4.4 horas, tornando-o bloqueado, pois esse ambiente determina que ele deve permanecer por pelo menos 55\% do tempo esperado. 


\subsection{Testes}
\label{subsec:testes}

Foram realizados testes do modelo proposto para identificar, observar, comparar e validar a evolução dos usuários. Para tal, foi realizada uma simulação de uso do ambiente utilizando a linguagem de programação Python. O programa consiste em um \textit{script} que vai modificando a data e hora do sistema e autentica usuários com diferentes características no ambiente.

Na implementação foi comparado todos os 31 dias do mês de janeiro de 2018 e faz quatro autenticações para cada usuário, sendo que todos entram às 8:00, permanecem até às 12:00 e entram novamente às 14:00, permanecendo até às 18:00. O horário que o usuário irá sair é definido por uma constante para cada usuário. Porém, para simular de maneira mais realista, essa constante é subtraída por um intervalo aleatório de até 1 hora, que possui 50\% de chances de ser aplicado. Na seção a seguir, são descritos os resultados obtidos.



\subsection{Resultados}
\label{subsec:resultados}

\begin{figure}[htbp]
	\centering
    \includegraphics[scale=.45]{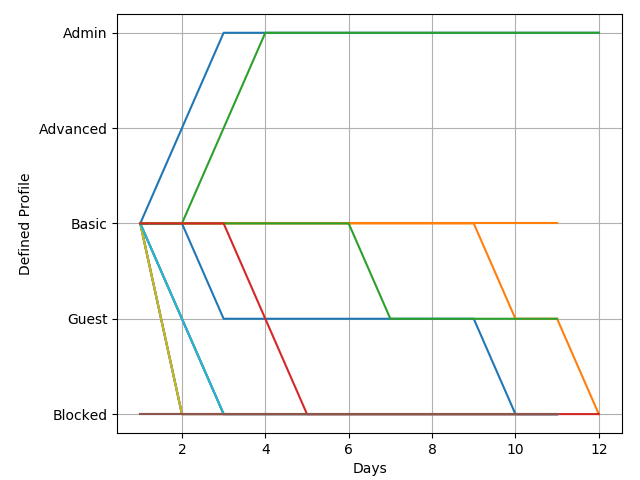}
    \caption{Evolução de perfis na Simulação.}
    \label{fig:simulacao}
\end{figure}

A Fig. \ref{fig:simulacao} apresenta os resultados obtidos na evolução dos usuários. Nele são mostrados os perfis dos usuários ao longo de doze dias. O eixo das ordenadas é o valor da variável qualitativa Perfil, a qual possui cinco possíveis estados: Bloqueado, Convidado, Básico, Avançado ou Administrador, e o eixo das abscissas significa a variação do tempo em dias. Cada linha desse gráfico representa um usuário diferente. Essa simulação foi feita com quinze usuários durante 31 dias, considerando o dia como o período de valência mais inferior. Nesse gráfico, não é possível distinguir usuários com características de frequência muito parecidas, pois suas linhas se sobrepõem, devido as atribuições individuais.

Porém, foi possível identificar que em 10,5\% das autenticações, o usuário obteve o perfil Bloqueado. Em 55,56\% das vezes, o usuário recebeu o perfil Convidado. Já em 10\% dos casos, o usuário obteve o perfil Básico. Os perfis Avançado e Administrador foram atribuídos em  22,78\% e 1,12\%, respectivamente. Esse resultado mostra que a ascensão de usuário básico para administrador não é fácil de ser obtida, já que o usuário deve possuir uma frequência quase que perfeita. Percebeu-se ainda uma tendência aos usuários terem seu perfil reduzido gradativamente e passarem a ser bloqueados em pouco tempo. Isso se deu por causa da distribuição das constantes dos usuários, as quais foram escolhidas de modo que todas ficassem bem distribuídas.

Com a simulação, foi possível perceber uma significativa melhora no tempo de resposta do sistema ao utilizar o perfil das últimas autenticações ao invés de sempre realizar uma requisição dos dados da esfera específica, como no trabalho anterior, diminuindo a sobrecarga de requisições sobre o servidor.


\section{Considerações Finais}
\label{sec:conclusoes}

Este artigo apresentou a implementação de um sistema para controle e gerenciamento para definição de perfis com base na privacidade dos dados em IoT.  Para tanto, fundamentou-se na definição de parâmetros e de suas características e funcionamento, que foram definidas em um modelo apresentado em esferas de ambientes e dados. A partir dos resultados obtidos nos testes foi possível identificar e definir diferentes perfis atribuídos a situações aleatórias. Também foi possível tratar a evolução e a redução da hierarquia com base em fatores que identificam a frequência de usuários nos ambientes testados. A evolução de perfis implementada demonstrou uma grande melhora na definição dos critérios de utilização e distribuição de parâmetros de privacidade para usuários. 

Neste trabalho, também foi possível simular evoluções nas permissões dos usuários dentro dos ambientes de acordo com variáveis genéricas. Exemplos de contribuições de controle como dia da semana, horário de chegada e saída, dias úteis, definição de horário para entrada em ambientes de acordo com perfis, entre outras regras que podem variar. Com isso, foi possível perceber que os períodos de valência escolhidos influenciam diretamente no tempo que as pessoas demoram para ter evoluções de perfis, trazendo vantagens e desvantagens. Por exemplo, se definimos como período de valência mais baixo o dia, um usuário que começa com perfil convidado, terá seu perfil evoluído para administrador em 3 dias, desde que possua uma frequência perfeita. Consequentemente, se escolhermos como perfil de valência mais baixo o mês, ele precisaria de 3 meses com frequência perfeita para evoluir seu perfil para administrador. Esses resultados obtidos contribuíram com a evolução de pesquisas realizadas anteriormente conforme citação no decorrer do trabalho, sendo possível utilizar um microcontrolador em substituição de arquitetura cliente servidor utilizadas anteriormente. Esse trabalho também contribui devido a utilização de dispositivos microcontroladores e protocolos de comunicação nos testes realizados.
 




\bibliographystyle{IEEEtran}
\bibliography{referencias}

\begin{thebibliography}{10}
\providecommand{\url}[1]{#1}
\csname url@samestyle\endcsname
\providecommand{\newblock}{\relax}
\providecommand{\bibinfo}[2]{#2}
\providecommand{\BIBentrySTDinterwordspacing}{\spaceskip=0pt\relax}
\providecommand{\BIBentryALTinterwordstretchfactor}{4}
\providecommand{\BIBentryALTinterwordspacing}{\spaceskip=\fontdimen2\font plus
\BIBentryALTinterwordstretchfactor\fontdimen3\font minus
  \fontdimen4\font\relax}
\providecommand{\BIBforeignlanguage}[2]{{%
\expandafter\ifx\csname l@#1\endcsname\relax
\typeout{** WARNING: IEEEtran.bst: No hyphenation pattern has been}%
\typeout{** loaded for the language `#1'. Using the pattern for}%
\typeout{** the default language instead.}%
\else
\language=\csname l@#1\endcsname
\fi
#2}}
\providecommand{\BIBdecl}{\relax}
\BIBdecl

\bibitem{Kubitza2016}
T.~Kubitza, A.~Voit, D.~Weber, and A.~Schmidt, ``{An IoT infrastructure for
  ubiquitous notifications in intelligent living environments},''
  \emph{Proceedings of the 2016 ACM International Joint Conference on Pervasive
  and Ubiquitous Computing Adjunct - UbiComp '16}, pp. 1536--1541, 2016,
  doi:10.1145/2968219.2968545.

\bibitem{wortmann2015internet}
F.~Wortmann and K.~Fl{\"u}chter, ``Internet of things,'' \emph{Business \&
  Information Systems Engineering}, vol.~57, no.~3, pp. 221--224, 2015.

\bibitem{gubbi2013internet}
J.~Gubbi, R.~Buyya, S.~Marusic, and M.~Palaniswami, ``Internet of things (iot):
  A vision, architectural elements, and future directions,'' \emph{Future
  generation computer systems}, vol.~29, no.~7, pp. 1645--1660, 2013.

\bibitem{razzaque2016middleware}
M.~A. Razzaque, M.~Milojevic-Jevric, A.~Palade, and S.~Clarke, ``Middleware for
  internet of things: a survey,'' \emph{IEEE Internet of Things Journal},
  vol.~3, no.~1, pp. 70--95, 2016.

\bibitem{banerjee2011everything}
P.~Banerjee, R.~Friedrich, C.~Bash, P.~Goldsack, B.~Huberman, J.~Manley,
  C.~Patel, P.~Ranganathan, and A.~Veitch, ``Everything as a service: Powering
  the new information economy,'' \emph{Computer}, vol.~44, no.~3, pp. 36--43,
  2011, doi: 10.1109/MC.2011.67.

\bibitem{li:18}
\BIBentryALTinterwordspacing
J.~Li, Q.~Yan, and V.~Chang, ``Internet of things: Security and privacy in a
  connected world,'' \emph{Future Generation Computer Systems}, vol.~78, pp.
  931 -- 932, 2018. [Online]. Available:
  \url{http://www.sciencedirect.com/science/article/pii/S0167739X17319817}
\BIBentrySTDinterwordspacing

\bibitem{leithardt:13}
V.~Leithardt, G.~Borges, A.~Rossetto, C.~Rolim, C.~Geyer, L.~Correia, D.~Nunes,
  and J.~Sá~Silva, ``A privacy taxonomy for the management of ubiquitous
  environments.'' 12 2013.

\bibitem{kraijak2015survey}
S.~Kraijak and P.~Tuwanut, ``A survey on iot architectures, protocols,
  applications, security, privacy, real-world implementation and future
  trends,'' 2015, doi:10.1049/cp.2015.0714.

\bibitem{ziegeldorf2014privacy}
J.~H. Ziegeldorf, O.~G. Morchon, and K.~Wehrle, ``Privacy in the internet of
  things: threats and challenges,'' \emph{Security and Communication Networks},
  vol.~7, no.~12, pp. 2728--2742, 2014.

\bibitem{sicari2015security}
S.~Sicari, A.~Rizzardi, L.~A. Grieco, and A.~Coen-Porisini, ``Security, privacy
  and trust in internet of things: The road ahead,'' \emph{Computer Networks},
  vol.~76, pp. 146--164, 2015.

\bibitem{zhang2014iot}
Z.-K. Zhang, M.~C.~Y. Cho, C.-W. Wang, C.-W. Hsu, C.-K. Chen, and S.~Shieh,
  ``Iot security: ongoing challenges and research opportunities,'' in
  \emph{Service-Oriented Computing and Applications (SOCA), 2014 IEEE 7th
  International Conference on}.\hskip 1em plus 0.5em minus 0.4em\relax IEEE,
  2014, pp. 230--234.

\bibitem{RUBIO2017205}
\BIBentryALTinterwordspacing
J.~E. Rubio, C.~Alcaraz, and J.~Lopez, ``Recommender system for
  privacy-preserving solutions in smart metering,'' \emph{Pervasive and Mobile
  Computing}, vol.~41, pp. 205 -- 218, 2017,
  doi:https://doi.org/10.1016/j.pmcj.2017.03.008. [Online]. Available:
  \url{http://www.sciencedirect.com/science/article/pii/S1574119217301645}
\BIBentrySTDinterwordspacing

\bibitem{ILATransactions}
G.~P.~R. Filho, L.~Y. Mano, A.~D.~B. Valejo, L.~A. Villas, and J.~Ueyama, ``A
  low-cost smart home automation to enhance decision-making based on fog
  computing and computational intelligence,'' \emph{IEEE Latin America
  Transactions}, vol.~16, no.~1, pp. 186--191, 2018.

\bibitem{hussein:17}
D.~Hussein, E.~Bertin, and V.~Frey, ``A community-driven access control
  approach in distributed iot environments,'' \emph{IEEE Communications
  Magazine}, vol.~55, pp. 146--153, 2017.

\bibitem{gusmeroli:13}
S.~Gusmeroli, S.~Piccione, and D.~Rotondi, ``A capability-based security
  approach to manage access control in the internet of things.''
  \emph{Mathematical and Computer Modelling}, vol.~58, no. 5-6, pp. 1189--1205,
  2013.

\bibitem{bernabe:16}
\BIBentryALTinterwordspacing
J.~Bernal~Bernabe, J.~L. Hernandez~Ramos, and A.~F. Skarmeta~Gomez, ``Taciot:
  Multidimensional trust-aware access control system for the internet of
  things,'' \emph{Soft Comput.}, vol.~20, no.~5, pp. 1763--1779, May 2016.
  [Online]. Available: \url{http://dx.doi.org/10.1007/s00500-015-1705-6}
\BIBentrySTDinterwordspacing

\bibitem{Rodrigues:17}
V.~F. Rodrigues, E.~Correa, C.~A. da~Costa, and R.~da~Rosa~Righi, ``On
  exploring proactive cloud elasticity for internet of things demands,'' in
  \emph{2017 XLIII Latin American Computer Conference (CLEI)}, Sept 2017, pp.
  1--10, doi:10.1109/CLEI.2017.8226417.

\bibitem{santos:17}
D.~Santos, J.~Cesconetto, J.~Martins, L.~Silva, I.~Ochôa, and V.~Leithardt,
  ``Ubipri pripro - controle e gerenciamento de perfis de usuários com base na
  privacidade de dados,'' in \emph{15 Escola Regional de Redes de
  Computadores}, ERRC 2017.\hskip 1em plus 0.5em minus 0.4em\relax SBC, 2017.

\bibitem{leithardt:18}
V.~Leithardt, L.~Henrique Andrade~Correia, G.~Borges, A.~Rossetto, C.~Rolim,
  C.~Geyer, and J.~M.~Sá~Silva, ``Mechanism for privacy management based on
  data history (ubipri-his),'' vol.~10, pp. 11--19, 03 2018, doi:
  10.5383/JUSPN.10.01.002.

\bibitem{7460923}
J.~H. Sarker and A.~M. Nahhas, ``Mobile rfid system in the presence of
  denial-of-service attacking signals,'' \emph{IEEE Transactions on Automation
  Science and Engineering}, vol.~14, no.~2, pp. 955--967, April 2017,
  doi:10.1109/TASE.2016.2547989.

\end{thebibliography}

\end{document}